\documentclass[twocolumn,english]{revtex4}
\usepackage[T1]{fontenc}
\usepackage[latin9]{inputenc}
\usepackage{graphicx}
\usepackage{amssymb}

\makeatletter

\usepackage{setspace}

\makeatletter

\makeatletter

\usepackage{esint}

\makeatother

\makeatother

\makeatother

\usepackage{babel}

\begin{document}

\title{Efficiency for preforming molecules from mixtures of light Fermi
and heavy Bose atoms in optical lattices: the strong-coupling-expansion
method}

\author{Anzi Hu$^{1}$, J. K. Freericks$^{2}$, M. M. Ma\'{s}ka$^{3}$, C.
J. Williams$^{1}$}

\address{$^{1}$Joint Quantum Institute, University of Maryland and National
Institute of Standards and Technology, Gaithersburg, Maryland 20899,
USA}

\address{$^{2}$ Department of Physics, Georgetown University, Washington,
D.C. 20057, USA }

\address{$^{3}$Department of Theoretical Physics, Institute of Physics, University
of Silesia, PL-40007 Katowice, Poland}

\begin{abstract}
We discuss the application of a strong-coupling expansion (perturbation
theory in the hopping) for studying light-Fermi-heavy-Bose (like $^{40}$K-$^{87}$Rb)
mixtures in optical lattices. We use the strong-coupling method to
evaluate the efficiency for pre-forming molecules, the entropy per
particle and the thermal fluctuations. We show that within the strong
interaction regime (and at high temperature), the strong-coupling
expansion is an economical way to study this problem. In some cases,
it remains valid even down to low temperatures. Because the computational
effort is minimal, the strong-coupling approach allows us to work
with much larger system sizes, where boundary effects can be eliminated,
which is particularly important at higher temperatures. Since the
strong-coupling approach is so efficient and accurate, it allows one
to rapidly scan through parameter space in order to optimize the pre-forming
of molecules on a lattice (by choosing the lattice depth and interspecies
attraction). Based on the strong-coupling calculations, we test the
thermometry scheme based on the fluctuation-dissipation theorem and
find the scheme gives accurate temperature estimation even at very
low temperature. We believe this approach and the calculation results
will be useful in the design of the next generation of experiments,
and will hopefully lead to the ability to form dipolar matter in the
quantum degenerate regime. 
\end{abstract}
\maketitle

\section{Introduction}

In recent years, there has been much interest in ultra-cold polar
molecules \cite{Carr}, as they have the promise for being a new
state of quantum degenerate matter, with unique properties. In order
to have a large dipole moment, the polar molecules must be in their
rovibrational ground state, where further cooling can ultimately lead
to quantum degenerate dipolar matter \cite{Jin_chem}. Such polar
molecules can have long-range, anisotropic or three-body interactions
\cite{key-5}, which may lead to novel quantum phases \cite{Santos,Pupillo}
and new applications in quantum information science \cite{DeMille}.
In most ultra-cold polar molecule experiments, one starts with a mixture
of ultra-cold gases of atoms of different species, for example various
isotopic combinations of K and Rb \cite{Minardi,Demille_o,Jin_Fesh,key-7,Bongs}.
These atoms can form a weakly bound state through a magnetic field
sweep over the Feshbach resonance \cite{grimm,Jin_Fesh}. To create
molecules with significantly higher dipolar moments, the loosely bound
Feshbach molecules are coherently transferred to a ground state with
very high efficiency through stimulated Raman adiabatic passage (STIRAP)
\cite{PJulienne,DJin,Jin_science,Carl}.

Although the rate of transferring a Feshbach molecule to the ground
state is very high, the overall efficiency for forming dipolar molecules
is still low due to the low efficiency of forming the loosely bound
Feshbach molecules during the field sweep. In Ref.~\cite{Jin_science},
the fermionic $^{40}$K and the bosonic$^{87}$Rb atoms are trapped
by an optical trap. The efficiency to form the Feshbach molecule depends
on the phase-space density of the two species. But, because the Fermi
cloud stops shrinking once it reaches the quantum degenerate regime,
and the Bose cloud continues to shrink as it Bose condenses, this
phase space density is low at low temperature, and never reaches appreciable
sizes at higher temperatures, as the clouds become more diffuse. On
the other hand, if the mixture is first loaded onto an optical lattice,
the motion of the atoms can be more strongly confined, and it is possible
to create a large area where exactly one atom of each species sits
at the same lattice site, leading to a reduced three body loss \cite{grimm}
and almost unit efficiency \cite{jim_eff} for pre-forming the molecules.

When mixtures of $^{40}$K and $^{87}$Rb are loaded into an optical
lattice, the atoms of each species are influenced by the optical lattice
differently\cite{LMathey}. With the same optical lattice depth,
the heavy atoms usually have much lower tunneling rate than the light
atoms because of their significantly larger mass. In Ref.~\cite{jim_eff},
it was shown that for sufficient lattice depths, the hopping rate
of Rb is more than an order of magnitude less than that of K. It is
therefore reasonable to ignore the quantum effects of the tunneling
of the heavy bosonic atoms while allowing the light fermionic atoms
to hop between nearest neighbors (a classical effect of the motion
of the Rb atoms is taken into account by averaging over all energetically
favorable distributions of Rb atoms). Such systems can be described
by the Fermi-Bose Falicov-Kimball model \cite{Falicov-Kimball,heavy-light-Fermi,Bose-Fermi_Iskin}.
Using this model, we quantitatively determine the probability of having
exactly one atom of each species per lattice site in order to optimize
the formation of dipolar molecules.

For the Falicov-Kimball model, the phenomena of pre-forming molecules
has been discussed for Fermi-Fermi mixtures or Fermi-hard-core-Bose
mixtures \cite{Gruber} on a homogeneous lattice and Fermi-Fermi
mixtures in a harmonic trap \cite{maciej_jim}. In previous work
\cite{jim_eff}, we considered the Fermi-soft-core-Bose mixtures
in a harmonic trap and determined the efficiency for pre-forming molecules
as the probability to have exactly one atom of each species per site.
We used inhomogeneous dynamical mean-field theory (IDMFT) and Monte
Carlo (MC) techniques to calculate the efficiency as well as the density
profile and the entropy per particle. Both of these methods have advantages
and disadvantages. The IDMFT approach is approximate for two-dimensional
systems, but it can calculate both the efficiency and the entropy
per particle. The MC method is numerically exact after it reaches
thermal equilibrium, but it can not calculate the contributions to
entropy coming from the heavy particles. Both methods require large
computational times to calculate properties of a trapped system of
reasonable size. Using these methods, we have shown that the efficiency
is significantly increased by first loading onto an optical lattice
before forming the molecules and near unit efficiency can be achieved
with parameters that are realistic for current experiments.

The efficiency of pre-formed molecules is also likely to be affected
by the heating (the temperature increase) induced by loading onto
an optical lattice \cite{Bloch_many_body_RMP_entropy,entropy_adiabaticloading_trey,temperature_Loadinglattice_entropy}.
Considering that thermal fluctuations generally destroy the ordering
and the localization of the particles, it is reasonable to expect
that the efficiency of having exactly one Rb atom and one K atom per
site should be reduced if the temperature becomes too high. On the
other hand, if the temperature is low enough, the presence of the
lattice significantly increases the efficiency, almost to unity in
the case of deep lattices. The temperature of the lattice system,
however, remains difficult to measure in experiment \cite{Thermo_Qi,Thermo_BH,thermo2_DeMarco_exp,Thermo4_nstat_Svistunov,themo1_troyer}.
Instead, it is often assumed that the process of loading atoms onto
optical lattices is adiabatic and therefore the total entropy of the
system is conserved \cite{temperature_Loadinglattice_entropy,adiabatic_loading,Bloch_many_body_RMP_entropy,entropy_adiabaticloading_trey,entropy_const}.
Determined based on the thermal properties of the gas before adding
lattices, the \emph{entropy per particle} is then used as an effective
temperature scale for the lattice system \cite{entropy_cooling_theory,Bloch_many_body_RMP_entropy}.
There are also several proposals for directly determining the temperature
for systems of bosonic atoms \cite{Numfluc_Temp_BH,onsite_numfluc_T_BH,TOF_SC_Therm_Pelster},
fermionic atoms \cite{Ketterle_nfluc_thermF} or the magnetic systems
\cite{Ketterle_Spin_gradient}. In Ref. \cite{Thermo_Qi}, a general
thermometry scheme is derived based on the fluctuation-dissipation
theorem. Through quantum MC simulation, this proposal is shown to
be applicable to the non-interacting fermionic systems \cite{Thermo_Qi}
and interacting bosonic systems \cite{Thermo_Qi_Troyer}.

In our paper, we discuss light-Fermi-heavy-Bose mixtures in optical
lattices based on the strong-coupling (SC) expansion method (perturbation
theory in the hopping). The calculation is oriented to develop an
efficient way of estimating the efficiency of pre-forming molecules
for a given experimental system. With the SC expansion method, we
obtain analytical expressions for the efficiency of pre-forming molecules,
the entropy per particle and the local charge compressibilities. The
behavior of the efficiency is studied both as a function of entropy
per particle and temperature. The determination of temperature is
further studied by applying the thermometry proposal in Ref. \cite{Thermo_Qi}
to the Fermi-Bose mixture. To benchmark the accuracy, we compare the
SC calculation with the IDMFT and MC calculations for all parameters
considered. Overall, we find excellent agreement between the three
methods. Such agreement even extends to the low temperature region
when the interaction is strong enough. This is particularly useful,
given the fact that the SC expansion calculation is significantly
faster than the IDMFT and MC calculations. Such a speedup makes it
possible to consider much larger lattice sizes to eliminate the boundary
effects, to scan the large parameter space for optimal parameter regions
for pre-formed molecules and to estimate the density fluctuations
and other properties.

The paper is organized as follows: in Sec. II, we discuss the Fermi-Bose
Falicov-Kimball model and define the efficiency for pre-forming molecules;
in Sec. III, we discuss the formalism for evaluating the efficiency,
the entropy, and other related quantities; in Sec. IV, we discuss
our result for various parameters and benchmark the SC expansion calculation
with the IDMFT and MC calculations; in Sec. V, we discuss the application
of the fluctuation-dissipation theorem for determining the temperature
and we present our conclusions in Sec. VI.

\section{Fermi-Bose Falicov-Kimball model}

For mixtures of heavy bosons and light fermions, such as $^{87}$Rb/$^{40}$K
mixtures, the hopping parameter for the heavy bosons ($^{87}$Rb)
is usually more than an order of magnitude less than the hopping parameter
for the light fermions ($^{40}$K) when one takes reasonable lattice
depths (greater than 15 Rb recoil energies) \cite{jim_eff}. In this
case, we can ignore the quantum-mechanical effects of the hopping
of the heavy bosons and describe such mixtures with the Fermi-Bose
Falicov Kimball model in the presence of a trap potential. The Hamiltonian
of this model is written as \begin{equation}
H=H_{0}+H_{h}=\sum_{j}H_{0j}+H_{h},\label{eq:H}\end{equation}
 with \begin{eqnarray}
H_{0j} & = & (V_{j}-\mu_{f})f_{j}^{\dagger}f_{j}+U_{bf}f_{j}^{\dagger}f_{j}b_{j}^{\dagger}b_{j}\nonumber \\
 & + & (V_{j}-\mu_{b})b_{j}^{\dagger}b_{j}+\frac{1}{2}U_{bb}b_{j}^{\dagger}b_{j}(b_{j}^{\dagger}b_{j}-1),\label{eq:H0}\end{eqnarray}
 and \begin{equation}
H_{h}=-\sum_{jj'}t_{jj'}^{}f_{j}^{\dagger}f_{j'}.\label{eq:hop}\end{equation}
 Here, $j$, $j'$ label the sites of a two-dimensional square lattice,
with a lattice constant, $a$. The symbols $f_{j}^{\dagger}$ and
$f_{j}$ denote the creation and annihilation operators for the fermions
at lattice site $j$, respectively. The symbols $b_{j}^{\dagger}$
and $b_{j}$ denote the creation and annihilation operators for the
bosons at lattice site $j$, respectively. The fermionic operators
satisfy the canonical anticommutation relation $\{f_{j},f_{j'}^{\dagger}\}=\delta_{j,j'}$
and the bosonic operators satisfy the canonical commutation relation
$[b_{j},b_{j'}^{\dagger}]=\delta_{j,j'}$. The quantity $V_{j}$ is
the trap potential, which is assumed to be a simple harmonic-oscillator
potential centered at the center of the finite lattice. We assume
that the $j$th site has the coordinate $(x_{j},y_{j})$, so that
$V_{j}$ can be written as \begin{equation}
V_{j}=t\left[\frac{\hbar\Omega}{2ta}\right]^{2}\left(x_{j}^{2}+y_{j}^{2}\right),\label{eq:trap}\end{equation}
 where $\Omega$ is the trap frequency. The quantity $\mu_{f}$ is
the chemical potential for fermions and $\mu_{b}$ is the chemical
potential for bosons. Combining the trap potential and the chemical
potentials, we can define an effective position dependent local chemical
potential for the fermions, $\mu_{f,j}=\mu_{f}-V_{j}$, and for the
bosons, $\mu_{b,j}=\mu_{b}-V_{j}$. $U_{bf}$ is the interaction energy
between fermions and bosons and $U_{bb}$ is the interaction energy
between the soft-core bosons. The symbol $-t_{jj'}$ is the hopping
energy for fermions to hop from site $j'$ to site $j$. We consider
a general $t_{jj'}$ for the formal developments in the earlier part
of the next section, but later specialize to the case of nearest-neighbor
hopping with amplitude $t$, which we will take to be the energy unit.
We also set the lattice constant, $a$ equal to one.

The efficiency for pre-forming molecules is defined as the averaged
joint probability of having exactly one boson and exactly one fermion
on a lattice site,\begin{equation}
\mathcal{E}=\frac{1}{N}\sum_{j}\langle\hat{P}_{1,1}^{j}\rangle=\frac{1}{N}\sum_{j}\mathrm{Tr}\left(\hat{P}_{1,1}^{j}e^{-\beta H}\right),\label{eq:efficiency_general}\end{equation}
with $\beta=(k_{B}T)^{-1}$ the inverse temperature. We define the
projection operator $\hat{P}_{1,1}^{j}$ for having exactly one boson
and one fermion at site $j,$\begin{equation}
\hat{P}_{1,1}^{j}=|n_{b,j}=1,n_{f,j}=1\rangle\langle n_{b,j}=1,n_{f,j}=1|,\label{eq:Proj_P}\end{equation}
 and $N$ is the smaller value in the total numbers of bosons and
fermions, $N_{b}$ and $N_{f}$. In our case, we assume equal number
of bosons and fermions, therefore $N=N_{b}=N_{f}$.

In general, one can obtain $\mathcal{E}$ directly from Eq. (\ref{eq:efficiency_general})
for a readily diagonalized Hamiltonian. In our case, the efficiency
$\mathcal{E}$ is derived by distinguishing the contribution from
terms corresponding to $n_{b,j}=1$ in the expression for the density
of fermions. We assume that the density of bosons and fermions at
site $j$, $\rho_{b,j}$ and $\rho_{f,j}$, can both be written as
a series in terms of the occupation number of bosons at site $j$
in the following way, \begin{equation}
\rho_{b,j}=\langle b_{j}^{\dagger}b_{j}\rangle=\sum_{n_{b,j}}\mathcal{W}_{j}(n_{b,j})n_{b,j},\label{eq:rhob}\end{equation}
 and \begin{equation}
\rho_{f,j}=\langle f_{j}^{\dagger}f_{j}\rangle=\sum_{n_{b,j}}\mathcal{W}_{j}(n_{b,j})\widetilde{n}_{f,j}(n_{b,j}).\label{eq:rhof}\end{equation}
 Here $n_{b,j}$ is the occupation number of bosons on site $j$,
$n_{b,j}=0,1,...$. The coefficient $\mathcal{W}_{j}(n_{b,j})$ is
the \emph{probability} of having exactly $n_{b,j}$ bosons at site
$j$ and the coefficient $\widetilde{n}_{f,j}(n_{b,j})$ is the probability
for having one fermion on site $j$ for the occupation number $n_{b,j}$.
The joint probability of having exactly one boson and one fermion
at site $j$ can be written as \begin{equation}
\mathcal{E}_{j}=\mathcal{W}_{j}(n_{b,j}=1)\widetilde{n}_{f,j}(n_{b,j}=1),\label{eq:effj}\end{equation}
 and the efficiency $\mathcal{E}$ is the average of $\mathcal{E}_{j}$
over all sites, \begin{equation}
\mathcal{E}=\frac{\sum_{j}\mathcal{E}_{j}}{N}=\frac{\sum_{j}\mathcal{W}_{j}(n_{b,j}=1)\widetilde{n}_{f,j}(n_{b,j}=1)}{N}.\label{eq:eff0}\end{equation}
 It can be shown that the expression for the efficiency obtained in
this way is the same as from Eq. (\ref{eq:efficiency_general}). Now,
the efficiency is obtained directly from the density of bosons and
fermions, which can be easily derived from the partition function
$\mathcal{Z}$ by taking derivatives with respect to the appropriate
chemical potentials, \begin{equation}
\rho_{b,j}=\frac{1}{\beta}\frac{\partial\mathrm{\ln}(\mathcal{Z})}{\partial\mu_{b,j}},\label{eq:nb0}\end{equation}
 and \begin{equation}
\rho_{f,j}=\frac{1}{\beta}\frac{\partial\mathrm{\ln}(\mathcal{Z})}{\partial\mu_{f,j}}.\label{eq:nf0}\end{equation}

To study the behavior of the efficiency as a function of the entropy
per particle, we evaluate the entropy per particle by dividing the
total entropy by the total number of particles, \begin{eqnarray}
s & = & S/(N_{b}+N_{f})\nonumber \\
 & = & \left(k_{B}\mathrm{\ln}(\mathcal{Z})-\beta k_{B}\frac{\partial\mathrm{\ln(\mathcal{Z})}}{\partial\beta}\right)/(N_{b}+N_{f}).\label{eq:s_first}\end{eqnarray}

It is worthwhile noticing that the formalism development in this section
is based on the grand-canonical ensemble. This ensemble is appropriate
because we assume that in the lattice system both the energy and the
number of particles fluctuate. This may seem in contradiction with
the use of the entropy per particle as an effective temperature scale,
because strictly speaking entropy is used as a parameter only for
the micro-canonical ensemble. This contradiction is resolved because
the entropy per particle is assumed to be conserved \emph{during}
the process of turning on the optical lattice. It is a conserved quantity
when comparing the systems before and after turning on the optical
lattice, which is particularly useful from the experimental point
of view, since the experiments often start without the optical lattices.
For the lattice system itself, assuming it is in thermal equilibrium,
it is more reasonable to consider it with the grand-canonical ensemble,
allowing the energy and number fluctuations. The difference between
the different ensembles of course is not problematic if we assume
the system is large enough to be in the thermodynamical limit, where
all three ensembles are equivalent.

\section{strong-coupling expansion formalism}

In this section, we explain the SC expansion formalism. We first discuss
the evaluation of the partition function $\mathcal{Z}$, approximated
by the second-order expansion in terms of the hopping, $H_{h}$. From
the partition function, we derive the expressions for the density
of fermions shown in Eqs.~(\ref{eq:nf}) to (\ref{eq:nfj2}), the
density of bosons in Eqs. (\ref{eq:nb}) to (\ref{eq:nb2}), the efficiency
in Eqs. (\ref{eq:Eff_total}) to (\ref{eq:eff2}) and the entropy
per particle in Eqs. (\ref{eq:s}) to (\ref{eq:S2}). For readers
who prefer to see the final expressions, we suggest skipping the following
derivation and referring to the equations listed above for the corresponding
quantities.

The evaluation of the partition function in the SC approach starts
with the exact solution of the atomic Hamiltonian $H_{0}$. Hence,
we use an interaction picture with respect to $H_{0}$, where for
any operator $\mathcal{A}$, we define the (imaginary) time-dependent
operator $\mathcal{A}(\tau)=e^{\tau H_{0}}\mathcal{A}e^{-\tau H_{0}}$.
The partition function is written using the standard relation, \begin{equation}
\mathcal{Z}=\mathrm{Tr}\left(e^{-\beta H}\right)=\mathrm{Tr}\left(e^{-\beta H_{0}}\mathcal{U}(\beta,0)\right).\label{eq:Z_def}\end{equation}
 Here, $\mathcal{U}(\beta,0)=\mathcal{T}_{\tau}\exp\left[\int_{0}^{\beta}H_{h}(\tau)d\tau\right]$
is the evolution operator with $\mathcal{T}_{\tau}$ being the time-ordering
operator for imaginary times. Expanding the exponential in $\mathcal{U}(\beta,0)$
up to second order in $H_{h}(\tau)$ and evaluating the resulting
traces with respect to equilibrium ensembles of $H_{0}$, we have
\begin{eqnarray}
\mathcal{U}(\beta,0) & \simeq & 1+\nonumber \\
 & + & \frac{1}{2}\int_{0}^{\beta}d\tau_{1}\int_{0}^{\beta}d\tau_{2}\mathcal{T_{\tau}}H_{h}(\tau_{1})H_{h}(\tau_{2}).\label{eq:U}\end{eqnarray}
Here, the first order correction to the partition function vanishes
because the hopping connects different sites. Substituting Eq.~(\ref{eq:U})
into Eq.~(\ref{eq:Z_def}), we can write the partition function as,
\begin{equation}
\mathcal{Z}=\mathcal{Z}^{(0)}(1+\mathcal{Z}^{(2)}),\label{eq:Z}\end{equation}
 where $\mathcal{Z}^{(0)}$ is the partition function in the atomic
limit ($t=0$), \begin{equation}
\mathcal{Z}^{(0)}=\mathrm{Tr}\left(e^{-\beta H_{0}}\right),\label{eq:}\end{equation}
 and $\mathcal{Z}^{(2)}$ corresponds to the second-order term in
the expansion of $\mathcal{U}$ divided by $\mathcal{Z}^{(0)}$, \begin{equation}
\mathcal{Z}^{(2)}=\frac{1}{2\mathcal{Z}^{(0)}}\mathrm{Tr}\left[e^{-\beta H_{0}}\int_{0}^{\beta}\int_{0}^{\beta}d\tau_{1}d\tau_{2}\mathcal{T}_{\tau}H_{h}(\tau_{1})H_{h}(\tau_{2})\right].\label{eq:Z2}\end{equation}

To simplify the notation, we introduce $\bar{\mu}_{f,j}(n_{b,j})$
to represent the negative of the fermionic part of the Hamiltonian
$H_{0j}$ {[}Eq.~(\ref{eq:H0})] when there is a fermion at site
j, \begin{equation}
\bar{\mu}_{f,j}(n_{b,j})\equiv\mu_{f}-V_{j}-U_{bf}n_{b,j},\label{eq:mubarf}\end{equation}
 and $\bar{\mu}_{b,j}(n_{b,j})$ for the negative of the bosonic part
of the Hamiltonian $H_{0j}$,\begin{equation}
\bar{\mu}_{b,j}(n_{b,j})\equiv(\mu_{b}-V_{j})n_{b,j}-U_{bb}n_{b,j}(n_{b,j}-1)/2.\label{eq:mubarb}\end{equation}
Both $\bar{\mu}_{f,j}$ and $\bar{\mu}_{b,j}$ depend on the number
of bosons at site $j$. The effective fugacities for bosonic and fermionic
particles can then be written as the exponential of $\bar{\mu}_{f,j}$
and $\bar{\mu}_{b,j}$ respectively,

\begin{equation}
\phi_{f,j}(n_{b})=\exp\left[\beta\bar{\mu}_{f,j}(n_{b,j})\right],\label{eq:w}\end{equation}
 and \begin{equation}
\phi_{b,j}(n_{b})=\exp\left[\beta\bar{\mu}_{b,j}(n_{b,j})\right].\label{eq:eps}\end{equation}
 The atomic partition function $\mathcal{Z}^{(0)}$ can then be written
in terms of the effective fugacities as, \begin{eqnarray}
\mathcal{Z}^{(0)} & = & \Pi_{j}\mathcal{Z}_{j}^{(0)},\label{eq:Z0}\end{eqnarray}
 where $\mathcal{Z}_{j}^{(0)}$ is the atomic partition function at
site $j$, \begin{equation}
\mathcal{Z}_{j}^{(0)}=\sum_{n_{b,j}}\phi_{b,j}(n_{b,j})(1+\phi_{f,j}(n_{b,j})).\label{eq:Z0j}\end{equation}

Now we evaluate the second term in the partition function, $\mathcal{Z}^{(2)}$
of Eq. (\ref{eq:Z2}). To satisfy the total number conservation, only
terms with $j=k'$ and $j'=k$ in $H_{h}(\tau_{1})H_{h}(\tau_{2})$
are non-zero after the trace and $\mathcal{Z}^{(2)}$ is reduced into
a sum of products of the fermionic annihilation and creation operators
at the same site, \begin{eqnarray}
\mathcal{Z}^{(2)} & = & \frac{1}{2}\int_{0}^{\beta}\int_{0}^{\beta}d\tau_{1}d\tau_{2}\sum_{jk}t_{jk}t_{kj}\nonumber \\
 &  & \times\mathrm{Tr}\left[\mathcal{T_{\tau}}e^{-\beta H_{0j}}f_{j}^{\dagger}(\tau_{1})f_{j}(\tau_{2})\right]/\mathcal{Z}_{j}^{(0)}\nonumber \\
 &  & \times\mathrm{Tr}\left[\mathcal{T}_{\tau}e^{-\beta H_{0k}}f_{k}(\tau_{1})f_{k}^{\dagger}(\tau_{2})\right]/\mathcal{Z}_{k}^{(0)}.\label{eq:Z2&Green}\end{eqnarray}
 Using the cyclic permutation relationship of the trace, the products
can be represented by the local atomic Green's function, \begin{equation}
G_{jj}(\tau)=-\mathrm{Tr}\left[\mathcal{T_{\tau}}e^{-\beta H_{0j}}f_{j}(\tau)f_{j}^{\dagger}(0)\right]/\mathcal{Z}_{j}^{(0)},\label{eq:G}\end{equation}
 and $\mathcal{Z}^{(2)}$ is expressed as integrations of the atomic
Green's functions in terms of their relative times, \begin{equation}
\mathcal{Z}^{(2)}=-\frac{1}{2}\int_{0}^{\beta}\int_{0}^{\beta}d\tau_{1}d\tau_{2}\sum_{jk}t_{jk}t_{kj}G_{kk}(\tau_{1}-\tau_{2})G_{jj}(\tau_{2}-\tau_{1}).\label{eq:Z2_G}\end{equation}

Solving the Heisenberg equation of motion for the annihilation operator
$f_{j}(\tau$), \begin{equation}
\frac{\partial f_{j}(\tau)}{\partial\tau}=e^{H_{0}\tau}\left[H_{0},f_{j}\right]e^{-H_{0}\tau}\label{eq:f_time_eqa}\end{equation}
 one easily finds the expression for the annihilation operator $f_{j}(\tau$)
in the interaction picture,\begin{equation}
f_{j}(\tau)=e^{\bar{\mu}_{f,j}(n_{b,j})\tau}f_{j}(0),\label{eq:f_tau}\end{equation}
 Substituting Eq.~(\ref{eq:f_tau}) into Eq.~(\ref{eq:G}), we obtain
the atomic Green's function in terms of the effective fugacities as,

\begin{equation}
G_{jj}(\tau)=\left\{ \begin{array}{cc}
-\sum_{n_{b}}\frac{\phi_{b,j}(n_{b})}{\mathcal{Z}_{j}^{(0)}}e^{\tau\bar{\mu}_{f,j}}, & \tau>0\\
\sum_{n_{b}}\frac{\phi_{b,j}(n_{b})\phi_{f,j}(n_{b})}{\mathcal{Z}_{j}^{(0)}}e^{\tau\bar{\mu}_{f,j}}. & \tau<0\end{array}\right.\label{eq:G_slvd}\end{equation}

We now perform the integration over $\tau_{1}$ and $\tau_{2}$ in
$\mathcal{Z}^{(2)}$ and obtain the final expression for $\mathcal{Z}^{(2)}$,\begin{widetext}
\begin{equation}
\mathcal{Z}^{(2)}=\frac{1}{2}\sum_{jk}t_{jk}t_{kj}\sum_{n_{b,j}n_{b,k}}\frac{\phi_{b,j}(n_{b,j})\phi_{b,k}(n_{b,k})}{\mathcal{Z}_{j}^{(0)}\mathcal{Z}_{k}^{(0)}}\frac{\beta\left[\phi_{f,j}(n_{b,j})-\phi_{f,k}(n_{b,k})\right]}{\bar{\mu}_{f,j}(n_{b,j})-\bar{\mu}_{f,k}(n_{b,j})}.\label{eq:Z2}\end{equation}
 \end{widetext} Note that the partition function we derived here
is not limited to the case of nearest-neighbor hopping with a uniform
hopping parameter $t$. Eq. (\ref{eq:Z2}) can be applied to describe
hopping between arbitrary sites $j$ and $k$ and the hopping parameter
$t_{jk}$ can vary over different sites of the lattice.

Observables are evaluated by taking appropriate derivatives of the
partition function. In calculating the derivatives, we truncate all
final expressions to include only terms through the order of $t_{jk}^{2}$.
Also note that because sites $j$ and $k$ are different sites, we
do not normally have denominators equal to zero in Eq. (\ref{eq:Z2}),
but in any case, the formulas are always finite as can be verified
by l'Hôpital's rule. During numerical calculations of the observables,
the denominator, $\bar{\mu}_{f,j}(n_{b,j})-\bar{\mu}_{f,k}(n_{b,j})$,
can become too small and cause numerical errors. In our calculations,
we use the Taylor expansion in terms of $\bar{\mu}_{f,j}(n_{b,j})-\bar{\mu}_{f,k}(n_{b,j})$
around zero when the absolute value of $\bar{\mu}_{f,j}(n_{b,j})-\bar{\mu}_{f,k}(n_{b,j})$
is less than $10^{-5}$.

The density distribution is evaluated by taking the derivative of
the partition function with respect to the appropriate local chemical
potential {[}Eqs.~(\ref{eq:nf0}) and (\ref{eq:nb0})]. For the density
of fermions at site $j$, the expression constitutes two terms corresponding
to derivatives from $\mathcal{Z}^{(0)}$ and $\mathcal{Z}^{(2)}$,\begin{eqnarray}
\rho_{f,j} & = & \frac{1}{\beta}\frac{\partial\mathrm{In}(\mathcal{Z})}{\partial\mu_{f,j}}=\rho_{f,j}^{(0)}+\rho_{f,j}^{(2)},\label{eq:nf}\end{eqnarray}
 where $\rho_{f,j}^{(0)}$ is the density of fermions in the atomic
limit, \begin{equation}
\rho_{f,j}^{(0)}=\frac{1}{\beta}\frac{\partial\mathrm{In}\left(\mathcal{Z}_{j}^{(0)}\right)}{\partial\mu_{f}}=\frac{\sum_{n_{b,j}}\phi_{f,j}(n_{b,j})\phi_{b,j}(n_{b,j})}{\mathcal{Z}_{j}^{(0)}},\label{eq:nfj0}\end{equation}
 and $\rho_{f,j}^{(2)}$ is the total contribution to the density
at site $j$ from particles hopping from all possible sites, \begin{eqnarray}
\rho_{f,j}^{(2)} & = & \sum_{k}t_{jk}t_{kj}\sum_{n_{b,j}n_{b,k}}\left\{ \frac{\phi_{b,j}(n_{b,j})\phi_{b,k}(n_{b,k})}{\mathcal{Z}_{j}^{(0)}\mathcal{Z}_{k}^{(0)}}\right.\nonumber \\
 & \times & \left[\beta\frac{(1-\rho_{j,k}^{(0)})\phi_{f,j}(n_{b,j})+\rho_{j,k}^{(0)}\phi_{f,k}(n_{b,k})}{\bar{\mu}_{f,j}(n_{b,j})-\bar{\mu}_{f,k}(n_{b,j})}\right.\nonumber \\
 & + & \left.\left.\frac{\phi_{f,k}(n_{b,j})-\phi_{f,j}(n_{b,k})}{\left[\bar{\mu}_{f,j}(n_{b,j})-\bar{\mu}_{f,k}(n_{b,j})\right]^{2}}\right]\right\} .\label{eq:nfj2}\end{eqnarray}
 Similarly, the density of bosons at site $j$ is written as a sum
of the atomic density and the hopping contribution as, \begin{eqnarray}
\rho_{b,j} & = & \frac{1}{\beta}\frac{\partial\ln(\mathcal{Z})}{\partial\mu_{b,j}}=\rho_{b,j}^{(0)}+\rho_{b,j}^{(2)},\label{eq:nb}\end{eqnarray}
 where \begin{equation}
\rho_{b,j}^{(0)}=\frac{1}{\beta}\frac{\partial\mathrm{In\left(\mathcal{Z}_{j}^{(0)}\right)}}{\partial\mu_{b,j}}=\frac{\sum_{n_{b,j}}n_{b,j}\phi_{b,j}(n_{b,j})\left[1+\phi_{f,j}(n_{b,j})\right]}{\mathcal{Z}_{j}^{(0)}},\label{eq:nbj0}\end{equation}
 and \begin{eqnarray}
\rho_{b,j}^{(2)} & = & \sum_{k}t_{jk}t_{kj}\sum_{n_{b,j}n_{b,k}}\left[(n_{b,j}-\rho_{b,j}^{(0)})\frac{\phi_{b,j}(n_{b,j})\phi_{b,k}(n_{b,k})}{\mathcal{Z}_{j}^{(0)}\mathcal{Z}_{k}^{(0)}}\right.\nonumber \\
 & \times & \left.\frac{\beta[\phi_{f,j}(n_{b,j})-\phi_{f,k}(n_{b,k})]}{\bar{\mu}_{f,j}(n_{b,j})-\bar{\mu}_{f,k}(n_{b,j})}\right].\label{eq:nb2}\end{eqnarray}

The expression for the efficiency is obtained from the density distributions
of the fermions and bosons. Similar to the expression for the densities,
the efficiency consists of two terms, one corresponding to the atomic
limit and one corresponding to the contributions from the hopping,
\begin{equation}
\mathcal{E}_{j}=\mathcal{E}_{j}^{(0)}+\mathcal{E}_{j}^{(2)},\label{eq:Eff_total}\end{equation}
 where\begin{equation}
\mathcal{E}_{j}^{(0)}=\frac{\phi_{b,j}(n_{b,j})\phi_{f,j}(n_{b,j})|_{n_{b,j=1}}}{\mathcal{Z}_{j}^{(0)}},\label{eq:Eff0}\end{equation}
 and \begin{eqnarray}
\mathcal{E}_{j}^{(2)} & = & \sum_{k}\sum_{n_{b,j},n_{b,k}}\frac{\phi_{b,j}(n_{b,j})\phi_{b,k}(n_{b,k})}{\mathcal{Z}_{j}^{(0)}\mathcal{Z}_{k}^{(0)}}\times\nonumber \\
 &  & \left\{ -\left(\frac{\phi_{b,j}(n'_{b,j})\phi_{f,j}(n'_{b,j})}{\mathcal{Z}_{j}^{(0)}}\right)_{n'_{b,j}=1}\right.\nonumber \\
 &  & \times\frac{\beta\left[\phi_{f,j}(n_{b,j})-\phi_{f,k}(n_{b,k})\right]}{\bar{\mu}_{f,j}(n_{b,j})-\bar{\mu}_{f,k}(n_{b,j})}\nonumber \\
 &  & +\delta_{n_{b,j},1}\left[\beta\frac{\phi_{f,j}(n_{b,j})}{\bar{\mu}_{f,j}(n_{b,j})-\bar{\mu}_{f,k}(n_{b,j})}\right.\nonumber \\
 &  & \left.\left.+\frac{\phi_{f,k}(n_{b,j})-\phi_{f,j}(n_{b,k})}{\left[\bar{\mu}_{f,j}(n_{b,j})-\bar{\mu}_{f,k}(n_{b,j})\right]^{2}}\right]\right\} .\label{eq:eff2}\end{eqnarray}

For the trapped system, the local chemical potential $\mu_{j}$ includes
both the global chemical potential $\mu$ and the trapping potential
$V_{j}$. The derivatives with regard to the local chemical potential
or the chemical potential leads to different physical quantities.
For the Fermi-Bose mixture considered here, the cross-derivatives
should also be evaluated. Specifically, the derivative with regard
to the global chemical potential $(\mu_{b}+\mu_{f}$) corresponds
to the total number fluctuations, \begin{eqnarray}
\kappa & = & \frac{\partial^{2}\ln\mathcal{Z}}{\partial^{2}(\mu_{b}+\mu_{f})}\nonumber \\
 & = & \beta\left[\langle\left(\hat{N}_{f}+\hat{N}_{b}\right)^{2}\rangle-\langle\hat{N}_{f}+\hat{N}_{b}\rangle^{2}\right].\label{eq:numfluc}\end{eqnarray}
Here we define the total number operators, $\hat{N}_{f}=\sum_{j}f_{j}^{\dagger}f_{j}$
and $\hat{N}_{b}=\sum_{j}b_{j}^{\dagger}b_{j}$. The \emph{global}
compressibility is introduced as the response of the local density
to the change of the global chemical potentials, \begin{eqnarray}
\kappa_{j}^{g} & = & \frac{\partial^{2}\ln\mathcal{Z}}{\partial(\mu_{f,j}+\mu_{b,j})\partial(\mu_{b}+\mu_{f})}\nonumber \\
 & = & \beta\left[\langle\left(f_{j}^{\dagger}f_{j}+b_{j}^{\dagger}b_{j}\right)\left(\hat{N_{f}}+\hat{N}_{b}\right)\rangle\right.\nonumber \\
 &  & \left.-\langle f_{j}^{\dagger}f_{j}+b_{j}^{\dagger}b_{j}\rangle\langle\hat{N_{f}}+\hat{N}_{b}\rangle\right].\label{eq:g_compress}\end{eqnarray}
 And the \emph{local} compressibility, or the onsite number fluctuation,
is determined from the derivatives with regard to the local chemical
potential,\begin{eqnarray}
\kappa_{j}^{l} & = & \frac{\partial^{2}\ln\mathcal{Z}}{\partial^{2}(\mu_{b,j}+\mu_{f,j})}\nonumber \\
 & = & \beta\left[\langle\left(f_{j}^{\dagger}f_{j}+b_{j}^{\dagger}b_{j}\right)^{2}\rangle-\langle f_{j}^{\dagger}f_{j}+b_{j}^{\dagger}b_{j}\rangle^{2}\right].\label{eq:local_compress}\end{eqnarray}
 Both the global and local compressibilities are derivatives of the
density distributions and can be obtained from the density expressions
above.

Finally, we obtain the expression for the entropy per particle defined
in Eq.~(\ref{eq:s_first}) by averaging the total entropy of the
system and we again write the entropy per particle in terms of the
atomic limit expression and the contributions from the hopping,\begin{eqnarray}
s & = & \frac{1}{N}\sum_{j}S_{j}^{(0)}+\frac{1}{N}\sum_{j}S_{j}^{(2)}.\label{eq:s}\end{eqnarray}
 Here $S_{j}^{(0)}$ is the entropy at site $j$ in the atomic limit,\begin{eqnarray}
S_{j}^{(0)}/k_{B} & = & \ln\left(\mathcal{Z}_{j}^{(0)}\right)-\beta\epsilon_{j},\label{eq:S0}\end{eqnarray}
 where the parameter $\epsilon_{j}$ corresponds to the onsite energy
at site $j$ in the atomic limit, \begin{eqnarray}
\epsilon_{j} & = & \frac{\partial\ln(\mathcal{Z}^{(0)})}{\partial\beta}\nonumber \\
 & = & \frac{1}{\mathcal{Z}_{j}^{(0)}}\sum_{n_{b,j}}\left\{ \bar{\mu}_{b,j}(n_{b,j})\phi_{b,j}(n_{b,j})\left[1+\phi_{f,j}(n_{b,j})\right]\right.\nonumber \\
 &  & +\left.\bar{\mu}_{f,j}(n_{b,j})\phi_{b,j}(n_{b,j})\phi_{f,j}(n_{b,j})\right\} .\label{eq:onsite_energy_e}\end{eqnarray}
 The averaged contributions from the hopping at site $j$ is $S_{j}^{(2)},$\begin{eqnarray}
S_{j}^{(2)}/k_{B} & = & \ln(1+\mathcal{Z}^{(2)})-\beta\frac{\partial\ln(1+\mathcal{Z}^{(2)})}{\partial\beta}\nonumber \\
 & = & -\frac{\beta^{2}}{2}\sum_{k}\sum_{n_{b,j}n_{b,k}}\frac{\phi_{b,j}(n_{b,j})\phi_{b,k}(n_{b,k})}{\mathcal{Z}_{j}^{(0)}\mathcal{Z}_{k}^{(0)}\left[\bar{\mu}_{f,j}(n_{b,j})-\bar{\mu}_{f,k}(n_{b,j})\right]}\nonumber \\
 &  & \times\left\{ \left[\phi_{f,j}(n_{b,j})-\phi_{f,k}(n_{b,k})\right]\right.\nonumber \\
 &  & \times\left[\bar{\mu}_{b,j}(n_{b,j})+\bar{\mu}_{b,k}(n_{b,k})-\epsilon_{j}-\epsilon_{k}\right]\nonumber \\
 &  & +\left.\bar{\mu}_{f,j}(n_{b,j})\phi_{f,j}(n_{b,j})-\bar{\mu}_{f,k}(n_{b,k})\phi_{f,k}(n_{b,j})\right\} .\label{eq:S2}\end{eqnarray}

This ends the discussion on the derivation of the SC expansion method
formulas. In general, the expressions obtained above are accurate
in the case when the hopping is much smaller than interaction strength
and the temperature is very high ($\beta t$ is small). In this parameter
region, the SC method can evaluate physical quantities, like the density
distribution, efficiency, compressibility and entropy, very efficiently.
The total number of particles is fixed by varying the chemical potentials,
$\mu_{b}$ and $\mu_{f}$. To maximize the efficiency and reduce three
body loss, we consider the low density region with attractive interspecies
interactions and repulsive bosonic interactions. For other strong-coupling
regions, the formulas developed above are equally applicable but not
further discussed in this paper.

\section{Results}

\subsection{Comparison with the IDMFT and MC calculations}

\begin{figure*}
\includegraphics[width=14cm]{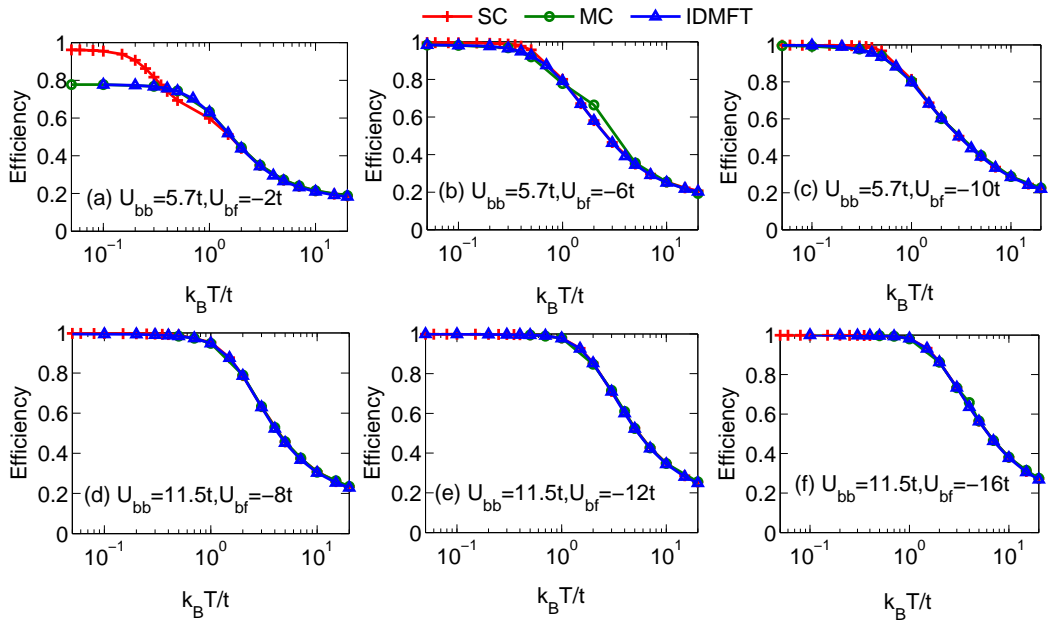}

\caption{\label{fig:Efficiency_VS_T}(Color on-line) Efficiency $\mathcal{E}$
as a function of temperature calculated by the SC (red cross), IDMFT
(blue triangle) and MC (green circle) methods. The interaction parameters,
$U_{bb}$ and $U_{bf}$, are shown in each plot. In (a), the SC calculation
differs from the other two methods for $T<1t/k_{B}$. For this region,
the SC expansion formulas derived here are no longer accurate. In
(b)-(f), all three methods give almost identical results. These calculations
also show that almost 100\% efficiency is reached for relatively strong
attraction, $U_{bf}\geq-6t$, at low temperature, $T<t/k_{B}$. }

\end{figure*}

For a perturbative method like the SC expansion method, it is always
necessary to determine the parameter regions where the approximation
is valid. Here, we use the previous results obtained from IDMFT and
MC methods \cite{jim_eff} as a reference to determine the accuracy
of the SC calculation. It is also worthwhile to notice that the three
methods require substantially different computational times. The SC
calculation usually takes less than 1 CPU hour while for the same
system the IDMFT calculation takes on the order of $10^{5}$ CPU hours.
We consider all the parameters used in the previous work \cite{jim_eff}.
The lattice is $50\times50$ square lattice with the trap frequency
$\Omega$ for both species fixed at $\hbar\Omega/2ta=1/11$. The parameters
$U_{bf}$ and $U_{bb}$ are chosen based on a typical experimental
setup : $U_{bf}/t=-8,\,-12,\,-16$ for $U_{bb}/t=11.5$ and $U_{bf}/t=-2,\,-6,\,-10$
for $U_{bb}=5.7$. The total number of bosons and fermions are set
to be 625. We consider the temperature range $0.05t/k_{B}$ to $20t/k_{B}.$

In Fig. \ref{fig:Efficiency_VS_T}, we show the efficiency as a function
temperature calculated with the three methods. Overall, we find excellent
agreement between the SC result and the IDMFT and MC calculations
and it is clear that high (unit) efficiency can be achieved when the
temperature is low ($T\sim0.1t/k_{B}$) and the interaction is large
compared with $t$. In the case of $U_{bf}=-2t$ and $U_{bb}=5.7t$,
the SC calculation starts to deviate greatly from the IDMFT and MC
calculation when $T\leq1t/k_{B}$. It is worth noting that for $T>1t/k_{B}$,
the SC calculations agree nicely with the other methods even for a
relatively weak cross-species interaction, $U_{bf}=-2t$.

The difference between the SC calculation and the other two methods
can be understood from the fact that the SC method is a perturbative
method based on the atomic limit of the Hamiltonian, $t=0$ and that
the properties derived from the SC expansion are dominated by the
atomic-limit behavior with relatively small corrections from the hopping.
In the atomic limit, bosons and fermions are completely localized
and the only density fluctuations are due to thermal fluctuations.
For the low density case considered here, the bosons always form a
plateau of unit filling at the center of trap at low temperature and
the fermions are attracted by the bosons one by one and form an almost
identical plateau. The efficiency therefore always converges to unity
as temperature deceases. In Fig. \ref{fig:Efficiency_VS_T}, we indeed
find the efficiency from the SC calculation always goes to one at
low temperatures. The convergence to unity is also true for the IDMFT
and MC calculations for all the parameters except for $U_{bf}/t=-2$
and $U_{bb}/t=5.7$. That's where the SC calculation differs from
the IDMFT and MC calculation. It is reasonable to assume that the
SC calculation can be applied to the region where the ground state
of the system is a localized, Mott insulator like state.

\begin{figure}
\includegraphics[width=7cm]{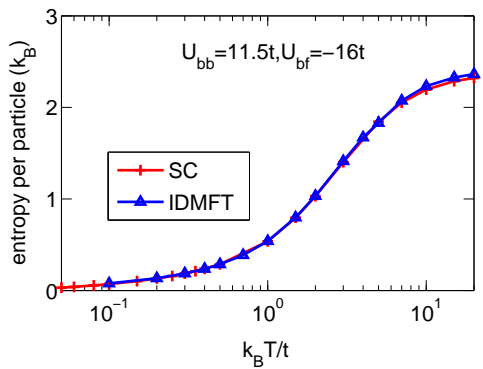}

\caption{\label{fig:EVST}(Color on-line) (a) Entropy per particle as a function
of temperature $T$. The SC calculation is marked with red crosses
and the IDMFT calculation by the blue line. We find excellent agreement
between the SC calculation and the IDMFT calculation. }

\end{figure}

\begin{figure}
\includegraphics[width=6.5cm]{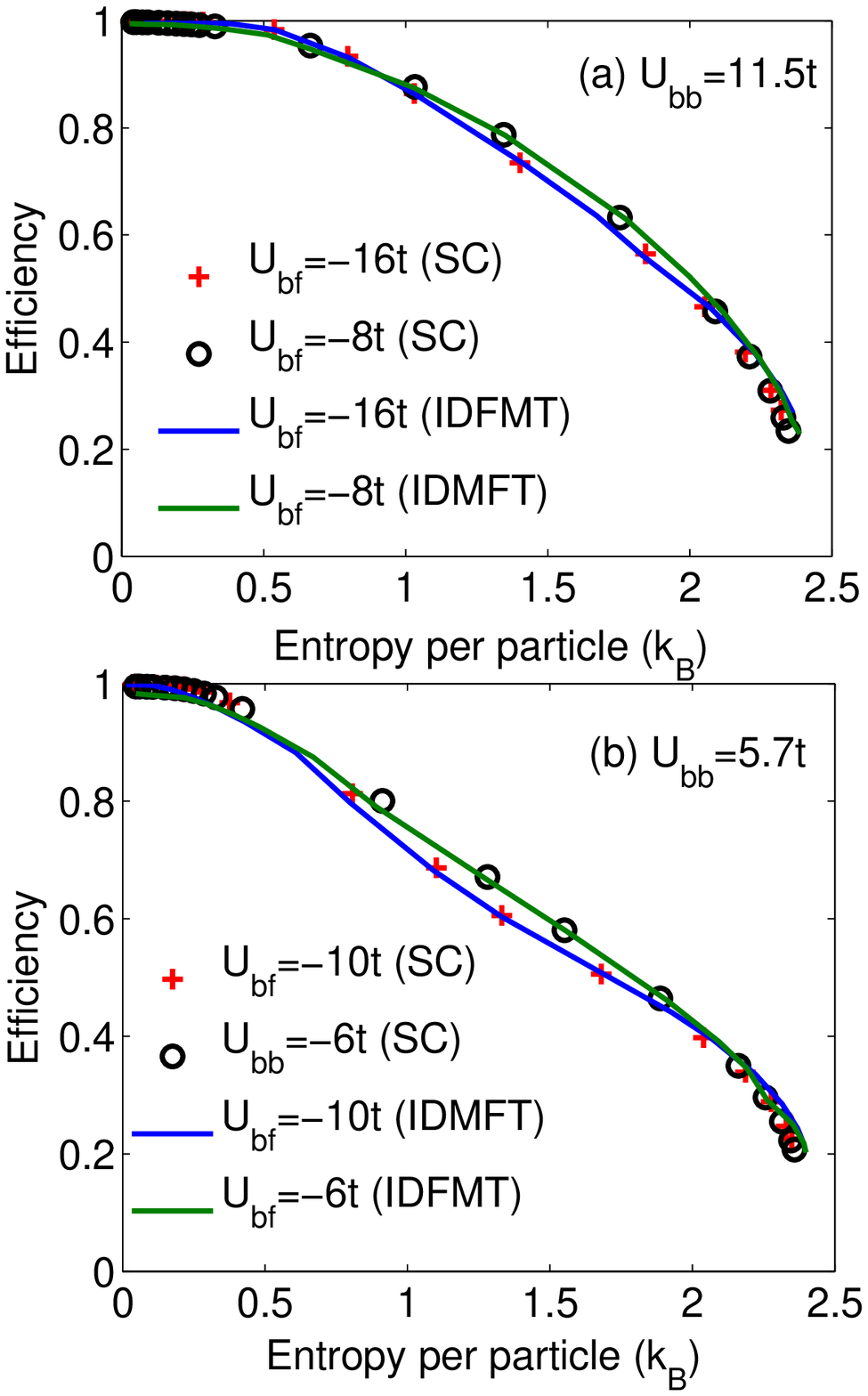}

\caption{\label{fig:effvsent_50}(Color on-line) Efficiency as a function of
entropy per particle for different interaction parameters. Note here
that we didn't include the case of $U_{bf}=-2t$, because it is already
shown in Fig. \ref{fig:Efficiency_VS_T} that the SC calculation is
not accurate for low temperatures in this case. In (a) and (b), we
consider two different bosonic interaction strengths and five different
inter-species interaction strengths. For all parameters, the efficiency
reaches $100\%$ when the entropy is very low. For an intermediate
entropy, with an entropy per particle around $1k_{B}$, the efficiency
is around $80\%$.}

\end{figure}

The SC calculation of the entropy per particle is also compared with
the IDMFT and MC calculations for all the parameters using Eqs. (\ref{eq:s}-\ref{eq:S2}).
The conclusion of the comparison is similar with the efficiency calculation,
that the SC calculation is accurate except for $U_{bf}=-2t$. In Fig.
\ref{fig:EVST} we use one example, $U_{bb}=11.5t$ and $U_{bf}=-16t$,
to represent all the cases where the SC calculation agrees with the
IDMFT calculation. As the temperature increases, the entropy per particle
starts to saturate at around $\sim2.3k_{B}$. In the next section,
we will show that this saturation is actually the result of finite-size
effects.

In Fig. \ref{fig:effvsent_50}, we show the behavior of the efficiency
as a function of the entropy per particle. This figure can be compared
with Fig. 2 in Ref. \cite{jim_eff}, where the IDMFT calculation
is discussed. We verify the findings from the previous work that for
strongly attractive inter-species interactions, an efficiency of $100\%$
can be achieved at low temperature (low entropy) region. For an entropy
per particle around $1k_{B}$, a $80\%$ efficiency can still be reached.
This efficiency is much higher than what has been achieved in experiment
\cite{Jin_science}.

In the following discussion on the SC calculation result, we no longer
consider the case of $U_{bf}=-2t$. This is also based on the consideration
that the interaction of $U_{bf}=-2t$ is too weak to achieve the desired
high efficiency of pre-formed molecules and therefore is not in the
parameter region of the main interest in this paper.

\subsection{Finite-size effects}

\begin{figure}
\includegraphics[width=7cm]{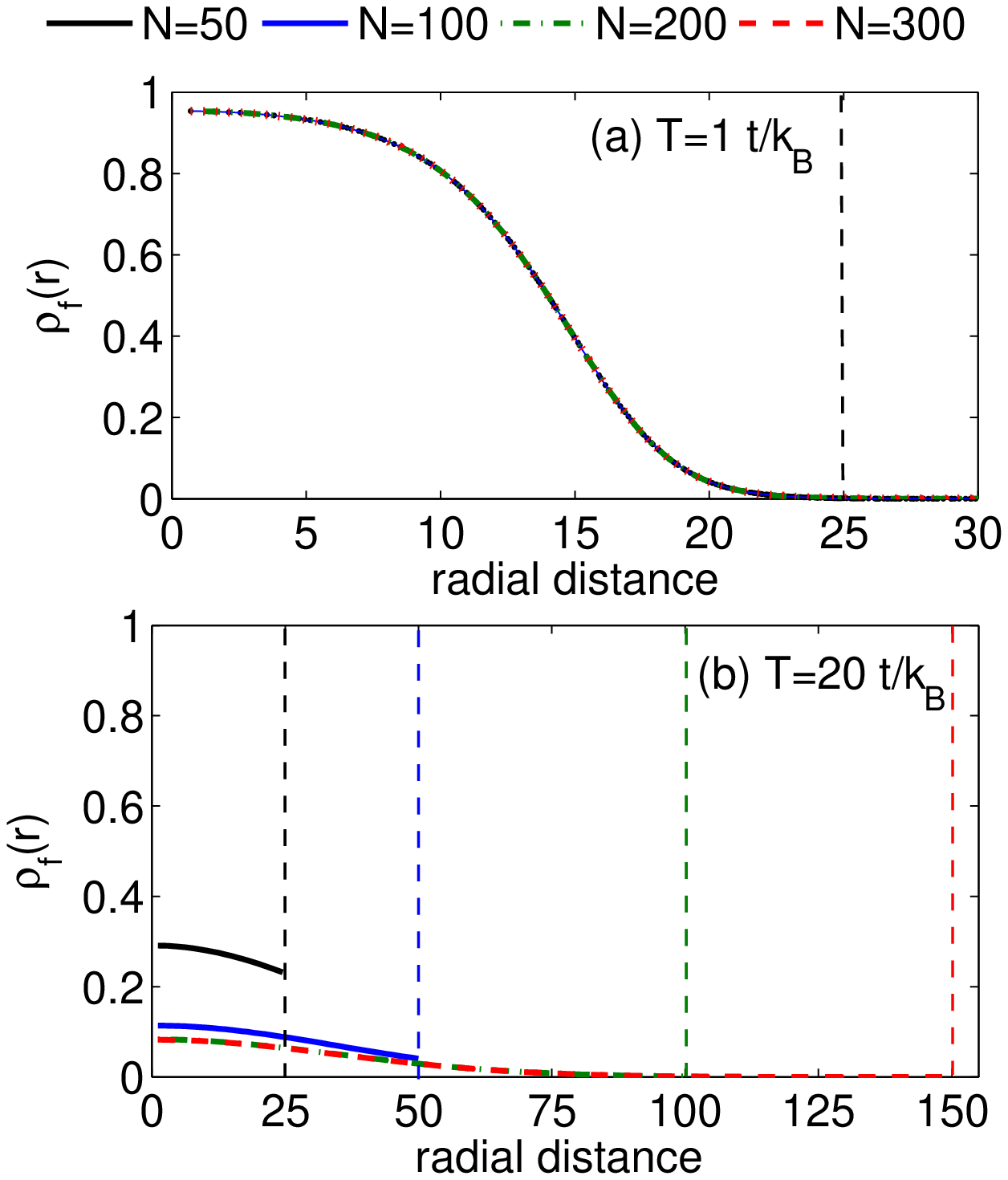}

\caption{\label{fig:finite-size}(Color on-line) Finite-size effects on the
radial density profile. We assume a two-dimensional $N\times N$ square
lattice with hard-wall boundary conditions. The dotted lines indicate
the boundaries of different lattices.The interaction parameters are
$U_{bf}=-16t$, $U_{bb}=11.5t$. We use the density distribution of
the fermionic particle to represent the general dependence of density
on the lattice size. In (a), we consider the case of low temperature,
$T=t/k_{B}$. Here, the density distribution is concentrated at the
center of the trap and there is no difference between different lattice
sizes. In (b), we consider the case of high temperature at $T=20t/k_{B}$.
Here, the density is confined mainly by the size of the lattice. For
$N=50$, the density is confined at the edge of the lattice, $r=25$.
For $N=100$, the density is again confined at the edge, $r=50$.
For both $N=200$ and $300$, the density goes to zero before reaching
the edge of the lattice and the two distributions overlap with each
other. We estimate that finite-size effects are eliminated for the
$300\times300$ square lattice for the trap frequency and number of
particles considered here.}

\end{figure}

\begin{figure}
\includegraphics[width=6.5cm]{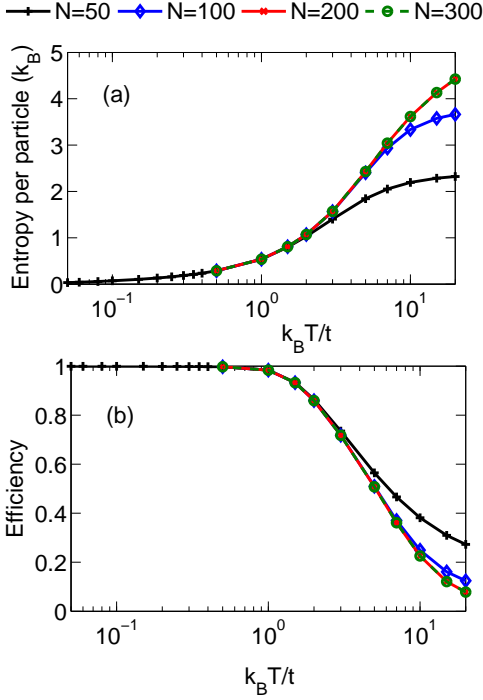}

\caption{\label{fig:ent&eff_finite}(Color on-line) Finite-size effects on
the entropy per particle and the efficiency. We assume a two-dimensional
$N\times N$ square lattice with hard-wall boundary conditions. The
interaction parameters are $U_{bf}=-16t$, $U_{bb}=11.5t$. In (a),
we show the behavior of the entropy per particle as a function of
temperature for different system sizes. We see the entropy is significantly
affected by the finite size when the lattice is smaller than around
$200\times200$. The finite-size effect is not noticeable at lower
temperature ($T<1t/k_{B}$).}

\end{figure}

\begin{figure}
\includegraphics[width=6.5cm]{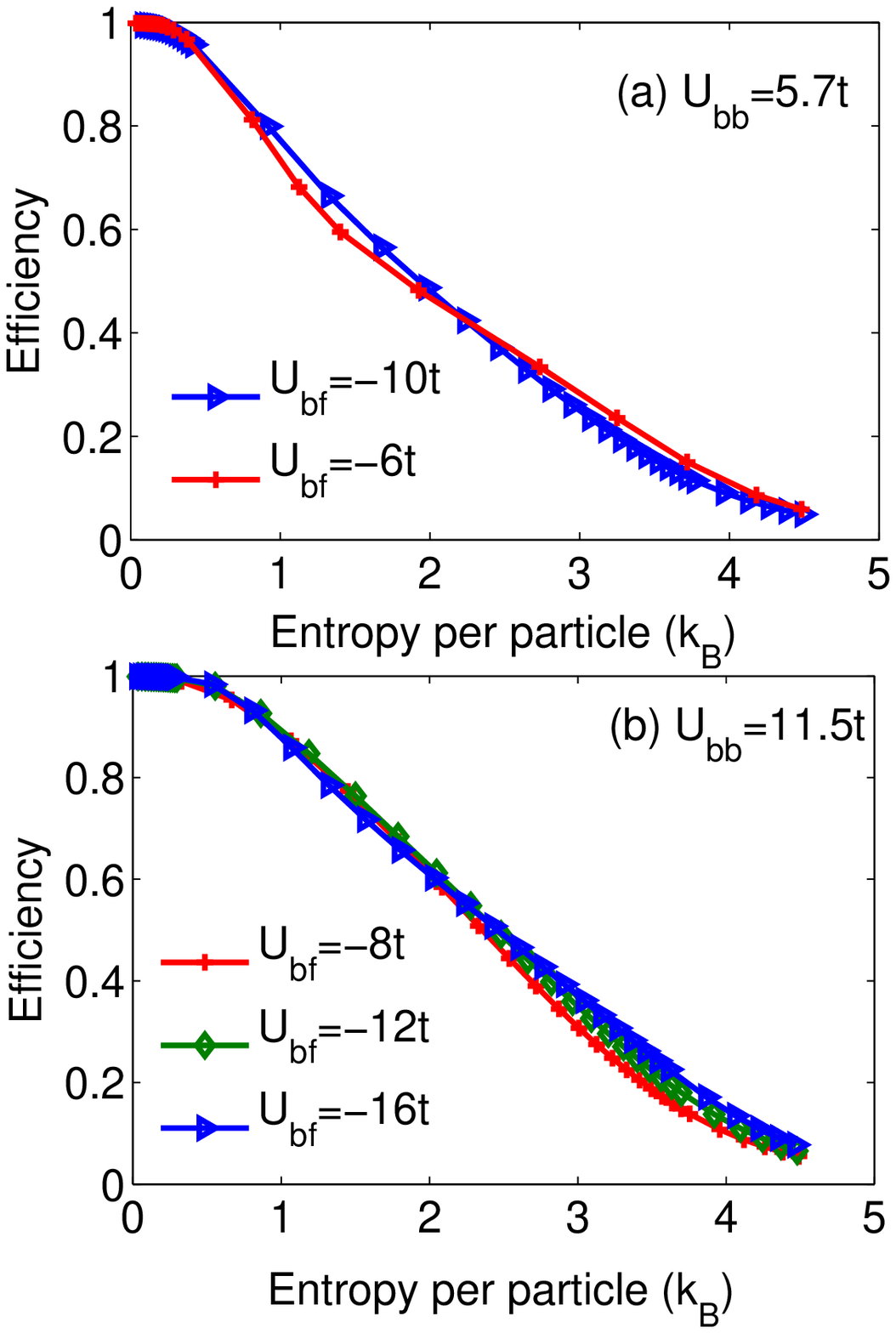}

\caption{\label{fig:Effiency-vs-T}(Color on-line) Efficiency as a function
of entropy per particle for a $300\times300$ square lattice system.
We consider 625 atoms for each species. Compared with Fig. \ref{fig:effvsent_50},
the efficiency is significantly higher for the same value of the entropy
per particle in the $300\times300$ lattice system when the entropy
per particle is large. On the other hand, the behavior is similar
in both lattice systems when the entropy per particle is less than
1$k_{B}$. The unit efficiency is reached roughly when the entropy
per particle is less than 0.5$k_{B}.$}

\end{figure}

In our calculations, we always assume a hard-wall boundary condition
at the edge of the lattice. In experiments, however, the atoms are
confined only by the trapping potential. This additional confinement
imposed by the boundary condition can potentially affect the accuracy
of our calculation. This finite-size effect can be neglected if the
system is so large that the atoms trapped by the trapping potential
almost never reach the edge the system. This, however, is not always
the case for the $50\times50$ lattice. This problem is difficult
to address with the IDMFT and MC methods, because of the high computational
costs. The SC method, on the other hand, can calculate much larger
systems for a fraction of the cost.

In this section, we discuss our calculation for different lattice
sizes and discuss finite-size effects for different lattice sizes.
To benchmark the SC calculations, the trap frequency and the total
number of particles are fixed for all different lattice sizes. We
assume the largest lattice sizes are sufficient to neglect the finite-size
effects. In Fig. \ref{fig:finite-size}, we show the density profile
as a function of the lattice size at two temperatures, $T=1t/k_{B}$
(a) and $T=20t/k_{B}$ (b). Here Fig. \ref{fig:finite-size} (a) represents
the scaling behavior in the low-temperature region, where there is
no significant difference between different lattice sizes and Fig.
\ref{fig:finite-size} (b) represents the scaling behavior in the
high-temperature region, where the system of small lattice size is
highly affected by the boundary effect. Note that the horizontal axes
are different scales in the two panels. The parameters used in the
plots are $U_{bf}=-16t$ and $U_{bb}=11.5t$. We find similar behavior
of the density profile for all the other parameters.

In Fig. \ref{fig:ent&eff_finite} (a), we show entropy per particle
as a function of temperature at different lattice sizes. In this plot,
we find that for small lattices, the entropy per particle becomes
\emph{saturated} at high temperature, while for large lattices it
keep increasing as the temperature increases. The saturation is understood
as the result of the finite-size effects. When the temperature is
high, atoms tend to expand to a larger area in the trap, which leads
to a large cloud size and higher entropy. When atoms expand to the
edge of the lattice, the possible occupied sites are now constrained
and the entropy stays similar even though the temperature increases,
hence the saturation. When the lattice is sufficiently large, atoms
can freely expand as the temperature increases and the entropy keeps
increasing.

The confinement of the atomic cloud in high temperature also affects
the efficiency calculation. In Fig. \ref{fig:ent&eff_finite} (b),
we find that the efficiency saturates to a higher value for smaller
lattices. This is because the confinement increases the density overlap
between the two species. In the low temperature region, the atoms
are close to unit filling at the center of the trap and the efficiency
is similar for all difference lattice sizes.

We find that a lattice of $300\times300$ sites is sufficient to eliminate
the finite-size effects for our parameter regions. Hence, we use this
lattice size for the efficiency and entropy per particle calculations.
In fig. \ref{fig:Effiency-vs-T}, we show the result for the efficiency
as a function of the entropy per particle. We estimate the calculation
result from the $50\times50$ lattice is accurate when the temperature
is around or below $T=1.25t/k_{B}$.

\section{Thermometry}

\subsection{Temperature and Density fluctuations}

Based on the fluctuation-dissipation theorem, the compressibility
can be related to the density fluctuations as \cite{Thermo_Qi},

\begin{equation}
\kappa=\frac{\partial\rho(r)}{\partial\mu}=\frac{1}{k_{B}T}\left[\langle\rho(r)N\rangle-\langle\rho(r)\rangle\langle N\rangle\right],\label{eq:n-c1}\end{equation}
where $\rho(r)$ is the radial density profile, $\mu$ is the chemical
potential and $N$ is the total number of particles. For a system
with a spherically symmetric harmonic trapping potential, $-V_{t}r^{2}$,
the local chemical potential at a radial distance $r$ is $\mu-V_{t}r^{2}$.
Within the local density approximation, the trapping potential is
interpreted as a variance in the chemical potential and the compressibility
in the trapped system can be re-written as a function of the density
gradient,

\begin{equation}
\frac{\partial\rho(r)}{\partial\mu}=-\frac{1}{2V_{t}r}\frac{\partial\rho(r)}{\partial r}.\label{eq:LDA}\end{equation}
These two equations lead to a simple relationship between the density
gradient and the density fluctuations in the trapped system,\begin{equation}
-\frac{1}{2V_{t}r}\frac{\partial\rho(r)}{\partial r}=\frac{1}{k_{B}T}\left[\langle\rho(r)N\rangle-\langle\rho(r)\rangle\langle N\rangle\right].\label{eq:therm_r1}\end{equation}
Based on this relationship, one can determine the temperature from
the independently measured density gradient and density fluctuations.
For a two dimensional system, a simplified relationship can be found
by integrating the above equation over all the two dimensional plane,

\begin{equation}
\frac{\pi}{V_{t}}\rho(0)=\frac{1}{k_{B}T}\left(\langle N^{2}\rangle-\langle N\rangle^{2}\right).\label{eq:sim_Them}\end{equation}
 Here, $\rho(0)$ stands for the density at the center of the trap.

With the development of \emph{in situ} measurements, it is now possible
to measure the density gradient and the fluctuations \cite{Ketterle_nfluc_thermF,chin}
in experiment and this thermometry scheme has shown promise to be
a reliable way of estimating the temperature \cite{Thermo_Qi,Thermo_Qi_Troyer}.
Here we test this method for the Bose-Fermi mixtures and Eqs. (\ref{eq:therm_r1}
and \ref{eq:sim_Them}) are extended to mixtures by considering the
density as the total density of both species and the total number
as the total number of both species. With the SC method, we calculate
the density gradient directly from the density profile expressions.
To simulate the fluctuations measured in the experiments, we use a
simplified MC simulation explained in the next section.

\begin{figure*}
\includegraphics[width=10cm]{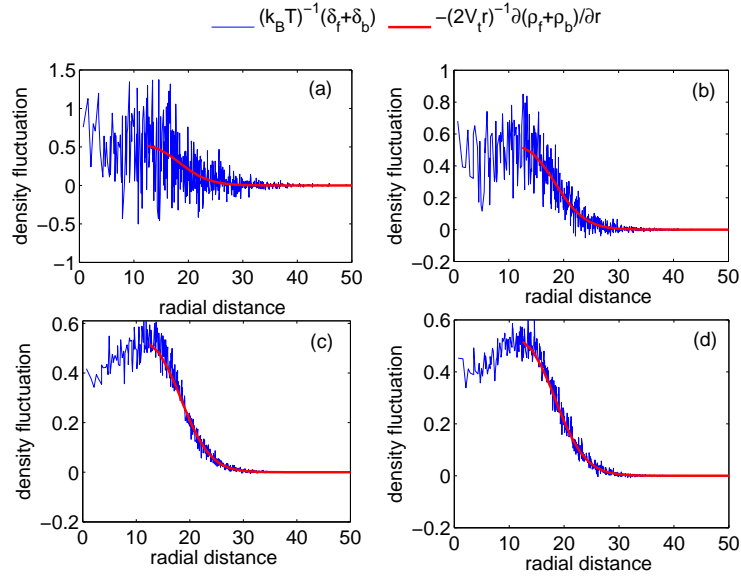}

\caption{\label{fig:thermo_1} (Color on-line) Density fluctuations averaged
over different numbers of samples. The density fluctuations shown
here are the total density fluctuations divided by the input temperature,
$T=2tk_{B}^{-1}$. All the fluctuations are compared with $-(2V_{t}r)^{-1}\partial(\rho_{b}+\rho_{f})/\partial r$.
According to Eq. (\ref{eq:therm_r1}), these two quantities should
be equal to each other. In (a)-(c), the total number of configuration
generated is $2\times10^{5}$, with a different sampling strategy.
In (a), one sample is taken at every $10^{3}$ configurations, which
gives a total of 200 samples to average over. The statistical error
in this case is very large. In (b), one sample is taken at every $100$
configurations, which gives a total of 2000 samples. The statistical
error is reduced compared with (a). In (c), the total number of samples
is $2\times10^{4}$. The statistical error is the smallest among (a)
to (c). In (d), a total of $2\times10^{6}$ configurations are generated
and $2\times10^{4}$ samples are taken at every 100 configurations. }

\end{figure*}

\subsection{Fluctuation calculation}

The MC simulation method generates a large collection of states (or
configurations) that satisfies the thermal equilibrium criteria. Such
collection of states constitutes a thermal ensemble. In the ensemble,
each state (or configuration) gives one density distribution, analogous
to one single shot image of the density in the experiment. By averaging
over all configurations, one obtains the averaged distribution of
particles. Deviations between different configurations are the fluctuations.
In our simplified MC method, we use the SC method to determine the
density distribution for a given temperature and then use the probability
as a reference for configuration generation. The ensemble of configurations
is decided to be large enough if it can reproduce the input probabilities. 

\emph{Determining the joint probability} : the joint probability,
$P_{n,m}^{j}$, is the joint probability of having $n$ bosons and
$m$ fermions at site $j$. For $m=1$, the joint probability of having
$n$ bosons and $1$ fermion at site $j$ can be found from the fermionic
density distribution, similar to the calculation of the local efficiency
$\mathcal{E}_{j}$ (indeed, $\mathcal{E}_{j}=P_{1,1}^{j}$),\begin{eqnarray}
P_{n,1}^{j} & = & P_{n,1}^{j(0)}-P_{n,1}^{j(1)}+P_{n,1}^{j(2)},\label{eq:Pjn1_overall}\end{eqnarray}
 where we again write the probability as a sum of the probability
in the atomic limit, \begin{eqnarray}
P_{n,1}^{j(0)} & = & \frac{\phi_{b,j}(n)\phi_{f,j}(n)}{\mathcal{Z}_{j}^{(0)}},\label{eq:Pjn10}\end{eqnarray}
 and the contributions from the hopping, \begin{eqnarray}
P_{n,1}^{j(1)} & = & \beta\sum_{k}\frac{\phi_{b,j}(n)\phi_{f,j}(n)}{\mathcal{Z}_{j}^{(0)}}\times\nonumber \\
 &  & \sum_{n_{b,j},n_{b,k}}\left[\frac{\phi_{b,j}(n_{b,j})\phi_{b,k}(n_{b,k})}{\mathcal{Z}_{j}^{(0)}\mathcal{Z}_{k}^{(0)}}\right.\nonumber \\
 &  & \times\left.\frac{\phi_{f,j}(n_{b,j})-\phi_{f,k}(n_{b,k})}{\bar{\mu}_{f,j}(n_{b,j})-\bar{\mu}_{f,k}(n_{b,k})}\right]\label{eq:Pjn11}\end{eqnarray}

\begin{eqnarray}
P_{n,1}^{j(2)} & = & \sum_{k}\sum_{n_{b,k}}\frac{\phi_{b,j}(n)\phi_{b,k}(n_{b,k})}{\mathcal{Z}_{j}^{(0)}\mathcal{Z}_{k}^{(0)}}\times\left[\frac{\beta\phi_{f,j}(n)}{\bar{\mu}_{f,j}(n)-\bar{\mu}_{f,k}(n_{b,k})}\right.\nonumber \\
 &  & +\left.\frac{\phi_{f,k}(n)-\phi_{f,j}(n_{b,k})}{\left[\bar{\mu}_{f,j}(n)-\bar{\mu}_{f,k}(n_{b,k})\right]^{2}}\right].\label{eq:Pjn12}\end{eqnarray}
 Once the joint probability $P_{n,1}^{j}$ is determined, the complementary
probability $P_{n,0}^{j}$ is found based on the relationship in the
atomic limit,

\begin{equation}
\sum_{n}\left[P_{n,1}^{j(0)}+\frac{\phi_{b,j}(n)}{\mathcal{Z}_{j}^{(0)}}\right]=1.\label{eq:P0_unit}\end{equation}
 Taking into the account the hopping contributions, we can write $P_{n,0}^{j}$
as,

\begin{eqnarray}
P_{n,0}^{j} & = & \frac{\phi_{b,j}(n)}{\mathcal{Z}_{j}^{(0)}}+P_{n,1}^{j(1)}-P_{n,1}^{j(2)}.\label{eq:Pjn0}\end{eqnarray}
We assume each lattice site is independent and the joint probability
at site $j$ is sufficient to determine the density distribution at
site $j$. The joint probabilities are evaluated for all the lattice
sites and stored in a table before the MC procedure. 

\emph{Simulation procedure}: we use a random number generator to generate
configurations with reference to the joint probability table. Specifically
the simulation includes the following steps:

1) Create a table for the values of $\widetilde{P}_{n,m}^{j}$ corresponding
to the sum of the joint probability of having \emph{up to} $n$ bosons
and up to $m$ fermions at site $j=1$, i.e. \begin{equation}
\widetilde{P}_{n,m}^{j}=\sum_{k=0}^{n}\sum_{l=0}^{m}P_{k,l}^{j}.\label{eq:}\end{equation}

2) Generate a random number $x$ between 0 and 1.

3) Find the smallest $\widetilde{P}_{n',m'}^{j}$ that is larger than
$x$. The number of bosons and fermions at site $j$ is then equal
to $n'$ and $m'$.

4) Repeat steps (2) and (3) to another site, $j=2$, until all the
lattice sites are considered. Store the configuration.

5) Repeat steps (2)-(4) $\mathcal{N}$ times to generate $\mathcal{N}$
configurations.

To avoid auto-correlation between adjacent configurations, we choose
every other $\mathcal{M}\gg1$ configurations as samples. The total
number of samples is then $N_{s}=\mathcal{N}/\mathcal{M}$. Averaging
over all the samples, we obtain the fermionic and bosonic part of
the density fluctuation as \begin{equation}
\delta_{f(b)}(r)=\langle\rho_{f(b)}(r)(N_{f}+N_{b})\rangle-\langle\rho_{f(b)}(r)\rangle\langle N_{f}+N_{b}\rangle,\label{eq:denfluc_fb}\end{equation}
 and the total density fluctuation is the sum of $\delta_{f}$ and
$\delta_{b}$. The total number fluctuation is defined as \begin{equation}
\Delta=\langle\left(N_{f}+N_{b}\right)^{2}\rangle-\langle N_{f}+N_{b}\rangle^{2}.\label{eq:nfluc_total}\end{equation}
Here the bracket stands for the averaging over all samples in analogy
to the experimental measurement of the fluctuations.

\begin{figure}
\includegraphics[width=6cm]{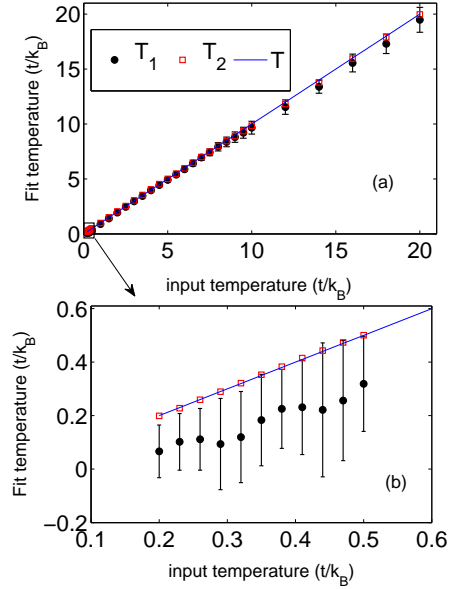}

\caption{\label{fig:thermo2}(Color on-line) Extracted temperature as a function
of the input temperature. The fluctuations are obtained from $2\times10^{4}$
samples out of $2\times10^{6}$ configurations. The value of $T_{1}$
is the mean of $T_{1}(r)$ averaged over $12<r/a<25$ and the error-bar
for $T_{1}$ is the standard deviation in $T_{1}(r)$ {[}Eq. (\ref{eq:T1_r})].
The value of $T_{2}$ is obtained through Eq. (\ref{eq:T2}). The
input temperature $T$ is drawn as a straight blue line in both plots.
In (a), we show our result for the full range of the input temperature,
from $T_{}=0.2t/k_{B}$ to $20t/k_{B}$. In this plot, we find very
good overall agreement of $T_{1}$ and $T_{2}$ with the input temperature
for the temperature range considered, particularly for $T>1t/k_{B}$.
In (b), we blow-up the area inside the black square in (a), which
corresponds to the low temperature region, where $T=0.2t/k_{B}$ to
$0.5t/k_{B}$. In this region, we find that $T_{1}$ shows large relative
fluctuations (deviation) from the mean value and the mean value of
$T_{1}$ differs relatively greater from $T$. The extracted temperature
$T_{2}$ however still shows excellent agreement with the input temperature. }

\end{figure}

\subsection{Results}

The fluctuation calculation is carried out for a $300\times300$ lattice
with all five sets of parameters. Overall we find very similar behavior
for all the parameters and we use parameters $U_{bb}=11.5t$ and $U_{bf}=-16t$
as an example. In our simulation, the fluctuations between different
configurations are from both the random number generator and the thermal
fluctuations. The difference between them is that the thermal fluctuations
are independent of ensemble sizes and the sampling size. We find that
the correct thermal fluctuation calculation requires a large number
of samples ($\sim10^{4}$) and large ensemble sizes ($\sim10^{6}$).
Because of the similarity between the simulation and experimental
measurement, this may suggest that a large number of shots are needed
in the experiments to obtain the correct thermal fluctuations. Note
that we consider here the results from a single plane as one shot,
not the averaged results over many planes as reported in Ref. \cite{Thermo_Qi_Troyer}. 

In Fig. \ref{fig:thermo_1}, we discuss the sampling effects by comparing
the fluctuations obtained from different samples with the compressibility
calculated from the density gradient $(2V_{t}r)^{-1}\partial\left[\rho_{f}(r)+\rho_{b}(r)\right]/\partial r$.
When the number of samples are small, the fluctuations are largely
random deviations from the average value. In Fig. \ref{fig:thermo_1}
(a), the fluctuations can be equally positive and negative, which
does not even satisfy the condition that the total fluctuations should
be always positive. As the number of samples grows, the random noise
starts to be averaged out and the fluctuations start to agree with
the fluctuation-dissipation theorem. In Fig. \ref{fig:thermo_1} (c),
the fluctuations agree very nicely with the relationship predicted
by Eq. (\ref{eq:therm_r1}). To show that $10^{4}$ samples are sufficient,
we consider an even larger ensemble, with $2\times10^{6}$ configurations
{[}Fig. \ref{fig:thermo_1}(d)] and find that the two ensembles produce
almost identical results. This shows that the fluctuation calculations
obtained in this way are independent of the ensemble size and should
correspond to the thermal fluctuations of the system.

With Eqs. (\ref{eq:therm_r1}) and (\ref{eq:sim_Them}), we define
two extracted temperatures, $T_{1}$ and $T_{2}$. Let $T_{1}(r)$
be the extracted temperature obtained in terms of the density fluctuations
and the density gradient,

\begin{equation}
k_{B}T_{1}'(r)=\frac{\delta_{f}(r)+\delta_{b}(r)}{(2V_{t}r)^{-1}\partial(\rho_{f}(r)+\rho_{b}(r))/\partial r},\,12<r/a<25.\label{eq:T1_r}\end{equation}
Here, we choose the radial distance to be larger than 12 lattice sites
because the quantity $(2V_{t}r)^{-1}\partial(\rho_{f}(r)+\rho_{b}(r))/\partial r$
diverges as $r\rightarrow0$ for a finite density gradient and for
small $r$, it goes to zero as the density develops a plateau at unit
filling at low temperature. The radial distance is less than 25 lattice
sites, because the density is almost zero in the outer regions and
that increases the relative error. Together, we find that $r$ between
12 and 25 sites to be the best region to fit the fluctuation and the
compressibility with each other. We also note that Eq. (\ref{eq:therm_r1})
still holds if one considers just the fermionic or bosonic part of
the system, i.e. keep only the index $f$ or $b$ in $\delta$ and
$\rho$. 

The temperature $T_{2}$ is obtained based on Eq. (\ref{eq:sim_Them}),
which translates into the following expression for the Bose-Fermi
mixture,

\begin{equation}
k_{B}T_{2}=\frac{\Delta}{\pi V_{t}^{-1}\left[\rho_{f}(0)+\rho_{b}(0)\right]}.\label{eq:T2}\end{equation}
Here $\rho_{f}(0)$ is the density of fermions at the center of the
trap and $\rho_{b}(0)$ the density of bosons at the center of the
trap.

In Fig. \ref{fig:thermo2}, we show $T_{1}$ and $T_{2}$ as a function
of input temperature. Overall, we find very good agreement between
$T_{1}$ and $T_{2}$ with the input temperature {[}Fig. \ref{fig:thermo2}
(a)]. We also find in the low temperature region, $T_{2}$ generally
fits better with the input temperature {[}Fig. \ref{fig:thermo2}
(b)]. This finding suggests that the statistical error introduced
by the numerically generated ensemble is lower in the calculation
of $T_{2}$ and this could be because the calculation of $T_{2}$
only involves the first-order observable, the density and the total
number fluctuation, whereas, for the calculation of $T_{1}$, we need
to calculate the second-order observable, the density fluctuation,
which may be more susceptible to statistical errors in the numerical
simulation.

\section{Conclusion}

The SC expansion method is a very efficient way of studying thermal
properties of strongly interacting systems. Through comparison with
the IDMFT and MC calculations, we show that the strong coupling expansion
method can be used for a wide range of parameters even at low temperature
when the attractive interaction between the two species is relatively
strong. We use the SC method to evaluate the finite-size effects in
our previous calculations. This leads to important modifications of
the efficiency and the entropy per particle at high temperature. The
SC calculation also provides a way to simulate experimental measurements
of the fluctuations. Based on the simulation, we find that the thermometry
proposal based on the fluctuation-dissipation theorem is accurate
for heavy-bose-light-fermi mixtures. This scheme suggests an effective
thermometry scheme that works in the extreme low temperature in the
deep lattice region. Overall, our work shows a promising way of creating
strongly interacting quantum degenerate dipolar matter by loading
the mixtures onto an optical lattice before the molecule formation.
In addition to higher efficiency, the molecules created in this way
are already situated in the optical lattice and can be directly adjusted
to realize the novel quantum phases that require the presence of a
lattice. It is also worth noting that the SC approach can be used
to study other mixtures with modifications. For Fermi-Fermi mixtures
like $^{6}$Li-$^{40}$K, it would require just truncating the heavy
bosonic states. For Bose-Bose mixtures like $^{87}$Rb-$^{133}$Cs
\cite{Nagerl}, the modification requires allowing for the superfluid
order to occur. 

\begin{acknowledgments}
J.K.F. was supported by a MURI grant from the AFOSR numbered FA9559-09-1-0617
and by the McDevitt endowment fund. Supercomputer time was provided
by the DOD HPCMP at the ARSC and ERDC computing centers. M.M.M. acknowledges
Grant No. NN 202 128 736 from Ministry of Science and Higher Education
(Poland).
\end{acknowledgments}

\end{document}